\documentstyle[twocolumn,aps]{revtex}
\begin{document}
\noindent{\bf Ziegler et al. reply:}  
In their Comment\cite{nt} on our recent Letter,\cite{ZHH}
Nersesyan and Tsvelik (NT) have raised questions
about the applicability of our
calculation of the density
of states (DOS) of a 2D d-wave superconductor (SC), though
they do not actually dispute the 
result of a nonzero Fermi Level DOS for a d-wave SC. 
They question the relevance of our exact
%explicit 
calculation of the DOS 
using a Lorentzian disorder distribution, claiming ``that models
with Lorentzian and Gaussian disorder belong to different universality
classes'', on grounds of a comparison with a straightforward perturbation
expansion in the disorder potential. We use a different method 
\cite{ZHH,ziegler1} to obtain lower bounds for the DOS for
other (e.g. Gaussian) distributions supporting our contention of a nonzero 
DOS at the Fermi level. NT challenge this method by alleging that
``similar discrepancies exist between the results obtained by this
method for another model of disorder - (2+1)D fermions with
random mass'', i.e. they contend that the method used in 
\cite{ziegler2} gives incorrect results.
%\indent 
\vskip .3cm
We have prepared a paper \cite{zhh2} which explicitly proves the existence 
of a nonzero lower bound for the Fermi level DOS for 
2D SC's with nodes (like p-wave, d-wave or extended s-wave SC's)
for a large class of disorder distributions, including the Gaussian.
The proof applies the work of Ref. \cite{ziegler1} to the case of 
interest. It also shows that the contribution of the "tails" of the
distribution are unimportant for the Fermi level DOS. Therefore,
our method of dealing with disorder in our Letter \cite{ZHH}
(using a Lorentzian distribution)
produces generic results for the physical system of a 2D SC with nodes,
with  disorder in the chemical potential.  
NT have not identified an error in our proof. 
%but rather state
%that the method Ref. \cite{ziegler1,ziegler2} 
%must be wrong because it produces incorrect
%results for other models.  
% This claim is based, however, on inconclusive numerical evidence, 
%as discussed below.
Their argument against it is based on inconclusive numerical evidence, 
as discussed below.
\vskip .01cm
\indent
The approach of NT \cite{NTW} approximates a lattice
system with
disorder in the chemical potential by a continuum model of Dirac
fermions with random gauge
fields. The more general model of Dirac fermions with all kinds of disorder 
allowed
(e.g. random mass) has been 
analyzed by Mudry et al. \cite{Mudry} who found that the critical
line corresponding to pure gauge field disorder is {\it unstable}
with respect to
the other kinds of disorder. They conclude that ``unless there exists a 
symmetry of the underlying lattice regularization'' forbidding other disorder
than random gauge fields the non--Gaussian cumulants are relevant 
perturbations. In short, the model considered by NT in Ref. \cite{NTW}
is most likely
not generic, since the physical system does not provide a symmetry
that allows the disorder to be described solely by random gauge fields.
In this regard we would like to emphasize that our work 
\cite{ZHH} is not ``aimed to debunk the entire
field theoretical approach to disordered systems'', but is  
actually in agreement with the conclusions 
of the no less field theoretical work by Mudry et al. \cite{Mudry}.
\vskip .3cm
\indent
{\bf 1) Comparison with perturbation theory}. 
It is clear that the Lorentzian
disorder distribution for which we obtained our exact results is quite 
special in terms of perturbation theory,
but we do not believe that for the Fermi level DOS it
will give qualitatively different results than
the impurity model with Gaussian disorder considered by NT. 
The fact that the Lorentzian disorder leaves the DOS of an s-wave SC 
unaffected shows that its tails 
do not necessarily lead to unphysical results.
Explicit calculations \cite{ziegler1,zhh2} show
that the tails of the distribution are in fact unimportant if the DOS
at the Fermi level is considered.
In any case, 
it is disingenuous to argue that because we can not reproduce 
a perturbative calculation with a Lorentzian distribution the result 
of our Letter must be "in a different universality class".
Since every moment of the distribution is infinite, 
perturbative arguments simply do not apply to this case.
%%%%%%%%%%%%%%%%%%%%%%%%%%%%%%%%%%%%%%%%%%%%%%%%%%%%%%%%%%%%%%%%%%
%Furthermore, even for cases where perturbation theory 
%converges
%%, e.g. Gaussian disorder,
%it is important to recall that the perturbation series controlled by the 
%width of the {\it distribution}
%$\gamma$ is a different expansion than the standard impurity
%perturbation series in the impurity {\it potential}.
%We see no reason to suppose that log--counting arguments carry over from
%one case to the other;
%NT's point that the perturbation series must contain the higher
%powers of $\log \Delta\over E$ is then far from obvious.\\
%Our result suggests that a proper resummation
%of the perturbation series, for those cases where it can be defined, would 
%show that the
%weak anomalies identified by NT do not drive the system to a new fixed
%point. \\
%
\vskip .3cm
\indent {\bf 2) Dirac fermions with random mass}.
In contrast to what NT imply, there is analytical
\cite{ziegler1,ziegler2} and numerical 
%\cite{hatsu}
\cite{hatsu,LeeWang}
evidence that Dirac fermions with random mass have a nonzero DOS at the
Fermi level.
For a  Gaussian disorder distribution with width $\gamma$
the DOS at the Fermi level was estimated in 
\cite{ziegler2} to be $\propto\exp(-\pi/\gamma)$, a nonanalytic behavior
with respect to 
the disorder that can not be obtained within perturbation theory. The related 
2D random bond Ising model is a highly controversial field of research
\cite{McCoy}.
Different groups have obtained different results, e.g. Refs.
11 and 12 of the Comment by NT disagree with their Ref. 8 on almost everything
but the specific heat. The numerical work quoted by NT as Ref. 10
agrees partially with their Refs. 8, 11 and 12 as well as with the result
of Ziegler \cite{ziegler2}. Because of the extreme weakness
of the $\log \log (T-T_c)$ divergence the numerical data are  not able
to resolve the question of divergence or finiteness of the specific heat;
the data can be fitted to both theories \cite{braak/ziegler}.
%Because the contribution to the DOS in the regime of interest is
%nonperturbative it is not surprising that methods based on perturbation
%theory (e.g. \cite{dotsenko}) seem to be in disagreement with rigorous
%and other nonperturbative treatments of this problem.
%Although we do not have a rigorous result for the specific heat
Therefore, there is no evidence that the methods of Refs. 
\cite{ziegler2} produce incorrect results.
%\indent To conclude, we have established 
%the finiteness of the DOS for 2D SC's with nodes by two 
%independent, nonperturbative and explicit methods \cite{ZHH,zhh2}.
\vskip .1cm
{\small 
K. Ziegler$^a$, M. H. Hettler$^{b,c}$, P. J. Hirschfeld$^b$
\vskip .01cm
\indent 
$^a$ M.P.I.--Phys. Kompl. Sys., D-70506 Stuttgart, Germany.
\vskip .01cm
\indent
$^b$ Dept. Physics, U. Florida, Gainesville, FL 32611, USA.
\vskip .01cm
\indent 
$^c$ U. Cincinnati, Mail Loc. 11, Cincinnati, OH 45221, USA.
\vskip .2cm
\noindent Received   March 1997
}
%\begin{thebibliography}{99}


\vskip -.6cm
\begin{references}
\vspace*{ -1.7cm}

\bibitem{nt} A. A. Nersesyan and A. M. Tsvelik, cond-mat/9701197.
\bibitem{ZHH}  K. Ziegler et al., Phys. Rev. Lett. 77, 3013 (1996). 
\bibitem{ziegler1} K. Ziegler, Nucl.Phys. B 285 (FS19), 606 (1987).
\bibitem{ziegler2} K. Ziegler, 
Nucl.Phys. B {\bf 344}, 499 (1990); Europhys.Lett. {\bf 14},
415 (1991).
\bibitem{zhh2} K. Ziegler et al., cond-mat/9703047.
\bibitem{NTW} A. A. Nersesyan et al., Phys. Rev.
Lett. {\bf 72}, 2628 (1994); Nucl. Phys. {\bf B 438}, 561 (1995).
\bibitem{Mudry} C. Mudry et al., Nucl. Phys. {\bf B 466},
382 (1996).
\bibitem{hatsu} Y. Hatsugai and P.A. Lee, Phys. Rev. {\bf B 48},
4204 (1993).
\bibitem{LeeWang} D.-H. Lee and Z. Wang, Phil.Mag.Lett. {\bf 73}, 145 (1996).
\bibitem{McCoy} B.M.  McCoy, 
%{\it "The connection between statistical 
%mechanics and quantum field theory"} 
in "Statistical Mechanics and Field
Theory", eds. V. V. Bazhanov and C. J. Burden, World Scientific, 
(Singapore 1995), 26-128.
%hepth/9403084. 
\bibitem{braak/ziegler}
D. Braak and K. Ziegler, Z.Phys. B {\bf 89}, 361 (1992).
%\bibitem{dotsenko}
%V. Dotsenko and V. Dotsenko, Adv.Phys. {\bf 32}, 129 (1983)
%\end{thebibliography}
\end{references}
\end{document}